\documentclass[useAMS,usenatbib]{mn2e}
\usepackage{graphicx}
\usepackage{times}
\usepackage{epsfig}
\usepackage{rotating}
\usepackage{sidecap}
\usepackage{longtable}
\usepackage{lscape}
\def\lesssim{\mathrel{\hbox{\rlap{\hbox{\lower4pt\hbox{$\sim$}}}\hbox{$<$}}}}
\let\la=\lesssim
\def\gtrsim{\mathrel{\hbox{\rlap{\hbox{\lower4pt\hbox{$\sim$}}}\hbox{$>$}}}}
\let\ga=\gtrsim

\def\fullsrc{IGR J17480-2446}
\def\ltsima{$\; \buildrel < \over \sim \;$}
\def\simlt{\lower.5ex\hbox{\ltsima}}
\def\gtsima{$\; \buildrel > \over \sim \;$}
\def\simgt{\lower.5ex\hbox{\gtsima}}

\title[IGR J17480-2446]{X-ray bursts and burst oscillations from the slowly
spinning X-ray pulsar IGR J17480-2446 (Terzan 5)}
\author[Motta et al.]{S.~Motta$^{1,2}$\thanks{E-mail: sara.motta@brera.inaf.it}, 
A. D'A\`i $^{3}$, A.~Papitto$^{4}$, A.~Riggio$^{4}$,  T.~Di Salvo$^{3}$, L.~Burderi$^{5}$, T.~Belloni$^{1}$ 
\newauthor L. Stella$^{6}$ and  R.~Iaria$^{3}$ \\ \\
$^{1}$INAF-Osservatorio Astronomico di Brera, Via E. Bianchi 46, I-23807   
Merate (LC), Italy\\
$^{2}$Universit\`a dell'Insubria, Via Valleggio 11, I-22100 Como,   
Italy\\
$^{3}$Dipartimento di Scienze Fisiche ed Astronomiche, Universit\`a di   
Palermo, via Archirafi 36, 90123 Palermo, Italy \\
$^{4}$INAF - Osservatorio astronomico di Cagliari, localit\'a Poggio dei Pini, strada 54, 09012 Capoterra, Italy\\
$^{5}$Dipartimento di Fisica, Universit\`a degli Studi di Cagliari, SP  Monserrato-Sestu, KM 0.7, 09042 Monserrato, Italy \\
$^{6}$INAF-Osservatorio Astronomico di Roma, Via di Frascati 33, I-00040 Monteporzio Catone (Roma)
}
\begin{document}
\maketitle
\begin{abstract}

 The newly discovered 11 Hz accreting pulsar, IGR J17480-2446,
located in the globular cluster Terzan 5, has shown several 
  bursts with a recurrence time as short as few minutes. The source shows the shortest recurrence time 
  ever observed from a neutron star. 
  Here we present a study
  of the morphological, spectral and temporal properties of 107 of the
  bursts observed by the Rossi X-ray Timing Explorer. The recurrence
  time and the fluence of the bursts clearly anticorrelate with the
  increase of the persistent X-ray flux. The ratio between the energy
  generated by the accretion of mass and that liberated during bursts
  indicate that Helium is ignited in a Hydrogen rich layer. Therefore we conclude that all the bursts shown by IGR J17480-2446 are Type-I X-Ray bursts. 

Pulsations could be detected in all the brightest bursts and no drifts
of the frequency are observed within 0.25 Hz of the spin frequency of
the neutron star. These are also phase locked with respect to the pulsations observed during the persistent emission and no rise of the rms associated to the pulse frequency is observed during the burst. This behavior would favor a scenario where the flash is ignited within a region which is consistent to be as large as the  neutron star surface.

\end{abstract}

\section{Introduction}

\begin{figure*}
\begin{center}
\includegraphics[width=18.0cm]{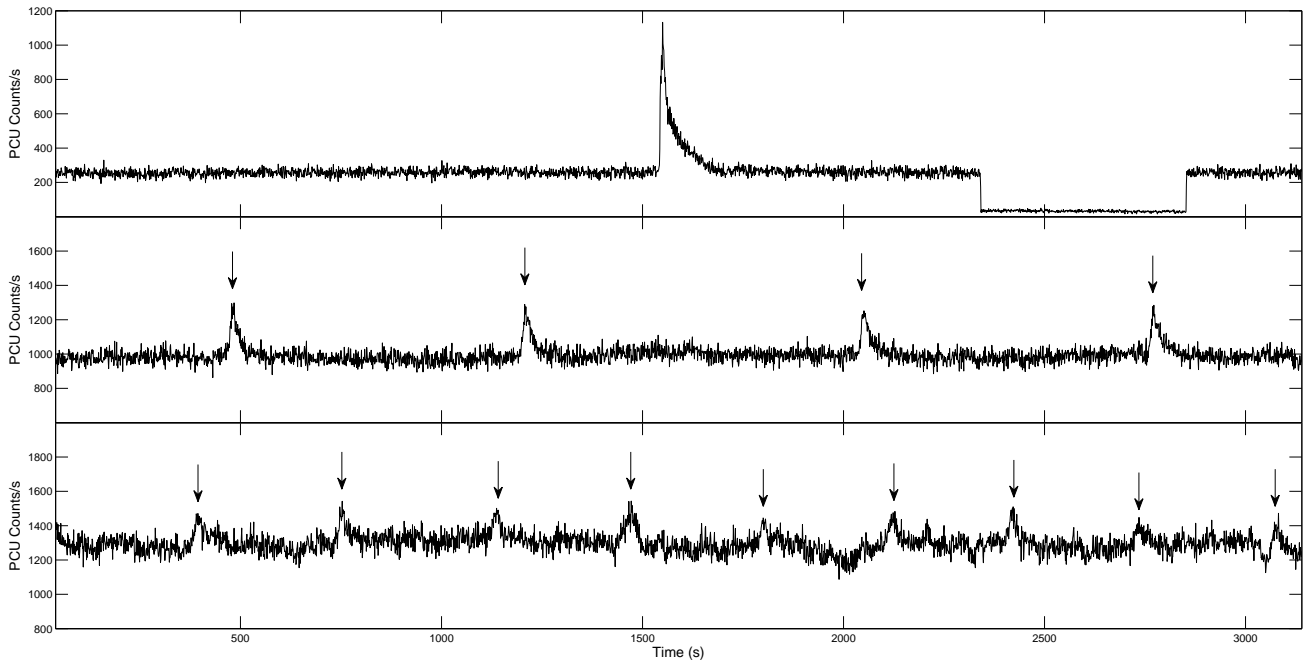}

\caption{Light curves of three RXTE observations of IGR J1748-2446. For each observation we plot the 1s-resolution  light curve from the beginning of the RXTE pointing. We show the first 3ks for each observation. From top panel: Obs. 95437-01-01-00, 95437-01-02-01, 95437-01-04-01. Arrows in the second and third panels from top mark the position of the bursts. Note that in Obs. 95437-01-01-00 a clear eclipse due to the moon occultation is visible (see \citealt{Strohmayer2010a}). The recurrence time between bursts is clearly different between the second and third observation (see text). }\label{fig:licu}
\end{center}
\end{figure*}

Type--I X-ray bursts result from unstable thermonuclear ignition of
accreted material on the surface of weakly magnetized neutron stars
(see  Lewin et al. 1983, \citealt{Strohmayer2006},
\citealt{Galloway2008},  and references therein for reviews on the
  subject). This material is accreted through Roche-lobe overflow
from a lower-mass companion star (low-mass X-ray binary, LMXB).  In
systems exhibiting bursts, the temperature and pressure at the base of
the accreted layer slowly increase until the nuclear energy generation
rate becomes more sensitive to temperature perturbations than to
radiative cooling.  At this point the resulting thermonuclear
instability leads to runaway burning of the matter that has been
deposited since the previous burst. During the flash, over 90\% of the
accreted hydrogen and helium is expected to burn into carbon and
heavier elements (\citealt{Woosley2004}).  For the next flash to
occur, a fresh layer of hydrogen/helium must first be accreted.

X-ray bursts are observed in about half of the total population of
LMXBs in the Galaxy, but only 15 sources from this sample show
coherent pulsations (Galloway et al. 2010) at, or within a few Hz
from, the spin frequency of the NS during the burst emission. To date,
only five LMXBs show coherent pulsations both during the burst and the
persistent X-ray emission (SAX J1808.4-3658, XTE J1814-338, HETE
J1900.1-2455 and IGR J17511-3057, and the peculiar case of Aql X-1
that showed coherent pulsations with an extremely low duty cycle, see
\citealt{Altamirano2010} and references therein). The spin frequency
in these sources is between 245 Hz and 550 Hz.

Notwithstanding the great amount of observational facts collected in
the last years, we still lack a clear understanding of the physical
mechanism responsible for the onset of the burst oscillations. It
would be important to understand why only $\sim$20\% of the bursting
X-ray sources show pulsations and why a drift of the order of a few Hz of the burst oscillation frequency is often observed during some type-I bursts (\citealt{Muno2002}), while in accreting millisecond pulsars this drift is less evident (e.g. Altamirano et al. 2010 and references therein).

The recent discovery of an 11 Hz accreting X-ray pulsar in the
Globular cluster Terzan 5 showing also burst oscillations at the
same frequency can greatly help to shed light on these questions,
definitely ruling out the need for high NS spin frequency as a
necessary ingredient both for the onset of X-ray burst and the burst
oscillations.

Furthermore, since the  burst peak flux from IGR J17480-2446  is exceptionally low, compared to the values observed in other bursting
sources, this source offers an excellent opportunity to study the effect of a hot NS contribution to the burst emission that arises when the burst luminosity is low compared to the persistent emission and in particular to the blackbody component (see \citealt{VP1986}). 


\section{Observations and data analysis}

{\it INTEGRAL} detected a transient source in the Globular Cluster Terzan 5
on 2010 October 10.365 (Bordas et al. 2010), tentatively attributed to
the known LMXB transient EXO 1745-248. However, follow-up Swift
observations refined the source position, excluding the association
with EXO 1745-248; the new source was therefore dubbed as IGR
J17480-2446. A {\it Chandra} observations confirmed the position of the
source (see \citealt{Pooley2010}) and its non-association with EXO
1745-248. The distance to Terzan 5 has been  estimated as 5.9$\pm$0.5 kpc (\citealt{Lanzoni2010}), which is the value we consider in the following.

Subsequent {\it Rossi X-ray Timing Explorer} (RXTE in the following)
observations of IGR J17480-2446 showed coherent pulsations at 11 Hz and the
presence of bursts and burst oscillations (\citealt{Altamirano2010a},
\citealt{Strohmayer2010}).  Timing analysis of the pulse period
revealed how the system has an orbital period of 21.327 hr and a
companion star mass between 0.4 and 1 M$_{\odot}$
(\citealt{Papitto2011}, \citealt{Strohmayer2010a}).

In this work, we focus on  the X-ray bursts
observed in the rising phase of the outburst of IGR J17480-2446
 with RXTE.  We consider observations from 13th to 17th
October (MJD 55482 to 55486, Obs. ID from 95437--01--01--00 to
95437--01--04--01).

\subsection{Outburst light curve, persistent emission and burst analysis}\label{Sec:Obs}

IGR J17480-2446 was observed daily by RXTE starting from MJD 55482.01
(Obs.ID from 95437-01-01-00, October 13th 2010), three days after its
discovery with INTEGRAL. A sample of the light curves observed with
RXTE is shown in Fig. \ref{fig:licu}. The count rate observed by the
PCU2 of the Proportional Counter Array (PCA) during the {\it
  persistent} emission\footnote{Even though the source is an X-ray
  transient, we refer to {\it persistent} emission as the main body of
  the outburst emission, and to {\it burst} emission when the source
  exhibits thermonuclear flashes} rises from $\sim250$ (October 13.0)
to $\sim1300$ c s$^{-1}$ PCU$^{-1}$ (October 16.7).

 To evaluate the X-ray {\it persistent} flux just prior each burst
onset, we extracted spectra only the PCU2 of the PCA
(3.5--25 keV) over a time interval of 32s before $T_{peak}$. We modelled the
spectrum with the sum of a blackbody and a Comptonized component,
\texttt{compps} (Poutanen \& Svensson 1996), and  computed
unabsorbed fluxes extrapolating in the 0.1-100 keV energy band. The source unabsorbed {\it persistent}
flux rises from $F_{pers}=0.54(5)\times 10^{-8}$ erg cm$^{-2}$
s$^{-1}$ on October 13.0, up to a peak value of $F_{pers}=1.7(2)\times
10^{-8}$ erg cm$^{-2}$ s$^{-1}$ during the last observation considered
here (October 16.7).



\begin{figure}
\begin{center}
\includegraphics[width=8.5cm]{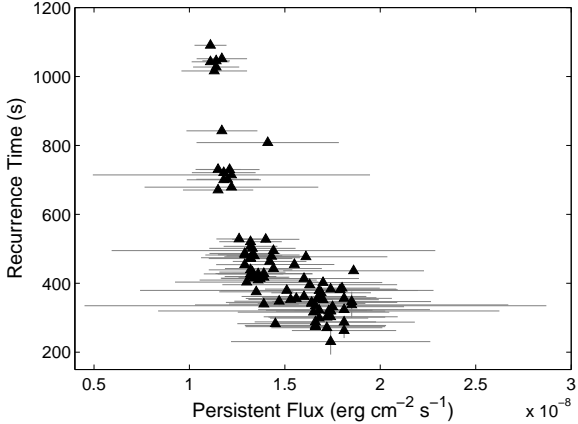}
\caption{Variation of the recurrence time as function of the {\it persistent} flux. }\label{fig:recurrence}
\end{center}
\end{figure}

X-ray bursts appearing at regular intervals are detected in all the RXTE observations we
consider (October 13th - 16th). The recurrence time between consecutive bursts decreases
from $\simgt24$ min to $\simeq5$ min while the {\it persistent} flux
increases (see Fig. \ref{fig:recurrence}). After October 17th, the bursts disappeared to appear again on October 18th. To analyze the morphological properties of the bursts observed during the
rising part of the outburst, we consider background-subtracted light curves extracted from
data taken in Event (122$\mu$s temporal resolution) and Good Xenon
(1$\mu$s temporal resolution) packing mode. Binning the light curves in 0.125s
intervals, we modelled the burst shape
with a five-parameter model: persistent count rate, start
time of the burst (T$_{rise}$), peak time (T$_{peak}$),
amplitude of the burst, exponential decay time of the burst
($\tau$). The model assumes a linear rise between T$_{rise}$ and
T$_{peak}$, and an exponential decay after the peak. The bursts show a
typical rise time between few and $\sim$20s and an exponential decay
time between $\sim$10 and $\sim$100s. As an example we show in the top panel of Fig. 
\ref{fig:licu} the shape of the first and most energetic burst observed by RXTE and in the following panels the typical shapes of the subsequent fainter bursts.

\begin{figure}
\begin{center}
\includegraphics[width=8.5cm]{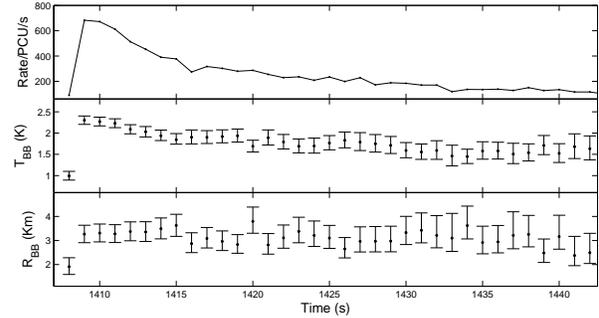}
\caption{Top panel: light curve of the strongest burst, observed on MJD 55482. Middle and bottom panel: evolution of radius and tamperature of the best-fit blackbody component. The integration time of the points is 2s at the peak and 4s during the last part of the decay.}\label{fig:profiles}
\end{center}
\end{figure}

To verify possible photospheric radius expansion episodes during the
bursts, we extracted PCU2 spectra of the
brightest burst, over time intervals 2 seconds long and modelled them with an absorbed
blackbody. The best-fit values of blackbody temperature and radius
are plotted in the middle and lower panel of Fig.\ref{fig:profiles},
respectively. The apparent radius observed at infinity has a constant value of $R_{app}^{\infty}=(3.3\pm0.5)\:d_{5.9}$ km, clearly indicating that no radius expansion takes place. The apparent radius has subsequently been converted into an effective radius taking into account the hardening factor and the effect of the gravitational redshift (see Sec. \ref{Sec:disc} for details).
We also measured the mean apparent radii for selected bursts following the first brightest one (see Tab. \ref{tab:radii}). To do this we followed Sztajno et al. (1986). For each selected burst we made a two-component fit to the combined persistent and burst emission, involving a black body component (associated to the NS emission) and a Comptonized component that represents the emission which is promptly radiated upon accretion of matter. We assumed that the latter component is not affected by the occurrence of a thermonuclear flash and we adopted the apparent blackbody radius associated to the other component as the mean radius of the region emitting the blackbody radiation during the Type-I X-Ray bursts. To better constrain the parameters of the persistent emission spectrum, we extracted the persistent emission spectrum immediately before the burst occurrence and we fitted it. Then we used this template model to fit the emission during the bursts, keeping the parameters of the comptonisation component and leaving the parameters of the blackbody component free to vary. The resulting measures (where the corrections adopted for the first burst have been applied) are listed in Tab. \ref{tab:radii}.

To verify our results, we attempted to stack data from several faint bursts observed at higher accretion rates (MJD 55485) and applied the same procedure followed for the single bursts. Analyzing the profiles of the fainter bursts, both rise and decay times often resulted different from one burst to another. For this reason, we selected bursts with similar profiles (i.e. similar rise and decay times) and averaged their spectra in order to obtain the measure of the mean radius and temperature of the blackbody emitting region. We obtained values consistent  with the ones reported in Tab. \ref{tab:radii}.\\


\begin{table} 
\renewcommand{\arraystretch}{1.3} 
\begin{center} 
\begin{tabular}{|c|c|c|c|} 

\hline 
No.	&		T						&	$R_{min}$	&	$R_{max}$	\\
	&		(keV)						&	(Km)	&	(Km)	\\
\hline 
2	&	$	1.41	_{-	0.08	} ^{+	0.03	}$	&	8.36	&	19.08	\\
6	&	$	1.34	_{-	0.08	} ^{+	0.07	}$	&	9.26	&	21.12	\\
10	&	$	1.36	_{-	0.10	} ^{+	0.08	}$	&	8.31	&	19.01	\\
28	&	$	1.27	_{-	0.06	} ^{+	0.04	}$	&	10.60	&	25.00	\\
31	&	$	1.37	_{-	0.06	} ^{+	0.03	}$	&	9.42	&	22.77	\\
32	&	$	1.28	_{-	0.06	} ^{+	0.06	}$	&	10.19	&	24.40	\\
80	&	$	1.17	_{-	0.01	} ^{+	0.05	}$	&	14.56	&	35.24	\\
93	&	$	1.17	_{-	0.03	} ^{+	0.01	}$	&	14.86	&	34.66	\\
106	&	$	1.17	_{-	0.03	} ^{+	0.03	}$	&	14.82	&	34.29	\\
\hline
\end{tabular}
\caption{Mean temperature and radius of the blackbody component associated to the thermonuclear burst emission. The upper and lower limit to the radius are corrected for the hardening factor f$_c$ and for gravitational redshift contribution.}\label{tab:radii} 
\end{center} 
\end{table} 

\begin{figure}
\begin{center}
\includegraphics[width=8.5cm]{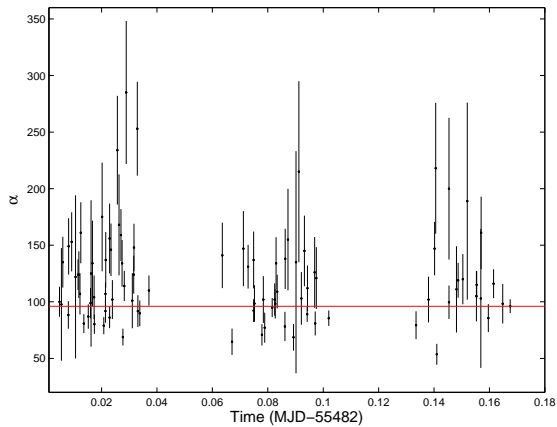}
\caption{Values of $\alpha$ obtained for all the consecutive bursts. The horizontal line marks the mean value $<\alpha>=96\pm3$.}\label{fig:alfa_grap}
\end{center}
\end{figure}

The energetics of each burst has been evaluated extracting a
PCU2 spectrum taken in a fast timing mode (Generic Event and Good
Xenon) over a time range containing the whole burst and of a variable
length depending on $\tau$ and on the variability of the {\it
    persistent} flux. We subtracted as background the spectrum of the persistent emission, extracted over an interval of 32 s before the onset of the burst (T$_{rise}$). We modelled  the
resulting spectrum with an absorbed blackbody. The fluence estimates
we obtain in the 0.1-100 keV band are listed in Table \ref{tab:burst}. The first burst is the
most energetic, showing a bolometric fluence
of $\mathcal{F}_{burst}=3.15(6)\times10^{-7}$ erg cm$^{-2}$. As the {\it
  persistent} flux of the source increases, bursts become more frequent
and less energetic as $\sim 3.5$d after the first observation they
have fluences of the order of $\sim 1\times10^{-8}$ erg cm$^{-2}$ (see Fig. \ref{fig:licu}).
We have then evaluated the ratio of the
  integrated persistent flux to the burst fluence,
$\alpha=F_{pers}t_{rec}/\mathcal{F}_{burst}$ for each burst for which the
recurrence time could be unambiguously defined. Modelling the
  observed values with a constant $\alpha$ we obtain an average value
  of $<\alpha>=96\pm3$, with a small variance of 1.32 over 89 points (see Fig. \ref{fig:alfa_grap}).
As we discuss in the following such values are compatible with a thermonuclear origin of the bursts (see also \citealt{Chakraborty2011}).
To secure such an identification, we searched for evidence of spectral softening during the tails of the bursts, which could be interpreted as cooling of the burst emission. While there is an indication of such a cooling during the first and brightest burst (see middle panel of Fig. \ref{fig:profiles}), a similar trend could not be observed for the subsequent, fainter bursts (see Galloway \& Zand 2010, \citealt{{Chakraborty2011}}).  We argue that this is not due to the absence of the
softening, but to the intrinsic difficulty in disentangling the persistent
and the burst emission when the tail of the burst is soon dominated by the
persistent emission.  Furthermore,  as pointed out by \cite{Sztajno1986} and \cite{VP1986}, the spectral analysis of X-ray bursts is systematically affected if the persistent emission contains a spectral component which originates from the outer layer of a hot NS, which also transfers an important energetic contribution to the burst emission. In this situation the net burst emission (i.e. excess above the persistent emission) is the difference between two blackbody spectra at different temperatures, corresponding to the total emission from the hot NS at different times. As a consequence, the net emission does not have a blackbody spectral distribution, therefore a \emph{standard} 
 blackbody spectral analysis on the net burst emission is not decisive in the identification of the burst nature. 


\subsection{Burst oscillations}

\begin{figure}
\begin{center}
\includegraphics[width=8.5cm]{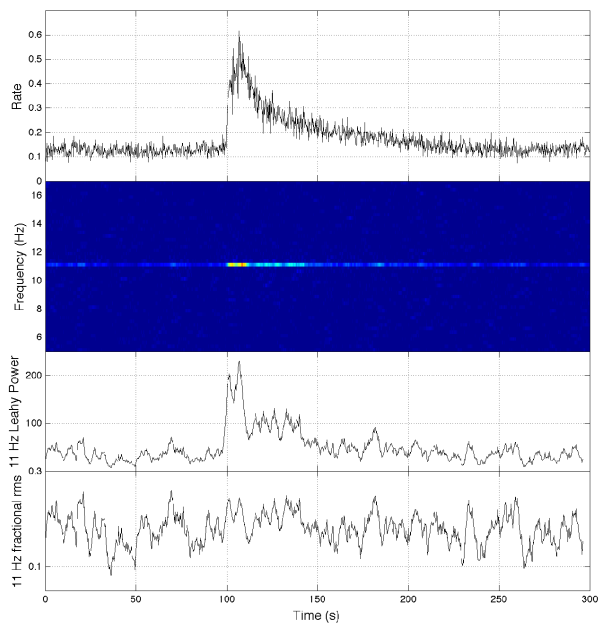}
\caption{Analysis of the burst observed on Oct. 13th. Top panel: light curve.  Second panel: Dynamical power density spectrum. Third panel: Evolution of Leahy power at 11.125 Hz. Bottom panel: corresponding fractional rms.}\label{fig:spectro}
\end{center}
\end{figure}

Burst-oscillation analysis was conducted using PCA data in Good Xenon
mode (for Obs. 95437-01-01-00) and Event Mode (for the following
observations), which provide full timing and spectral information.
We produced a dynamical power density spectrum (DPDS) computing Fast
Fourier Transforms (FFTs) of overlapping windows of data of length 4s,
stepped by 0.25s (Fig.\ref{fig:spectro}, second panel from top). As an
example we show the DPDS for the case of the first burst
observed on October 13th. In the third
panel from top of Fig. \ref{fig:spectro} we show the evolution of the
11 Hz Leahy power (\citealt{Leahy1983}) as function of time for that burst. Here the rise in
the pulsation power during the burst is driven by the increase in
count rate.  The increase is consistent with a constant fractional rms, as shown in the bottom panel of Fig. \ref{fig:spectro}.  We note that no drifting of the pulsation frequency
is observed within 0.25 Hz from the best-fitting spin
period. Interestingly, while the decay phase of the burst follows the
expected exponential profile, the evolution of the power (see
Fig. \ref{fig:spectro}, third panel from the top) clearly shows a multi-peak structure that does
not reflect the decay in the light curve. A similar case in observed in XTE J1814-338 (\citealt{Strohmayer2003}).

The other bursts also show an increase (although statistically less evident) in
the pulsation power. In particular, we observed a stronger power
in the 11 Hz pulsations in bursts taking place during 
MJD 55482 and 55483.
Later in the outburst, the increase in power becomes difficult to observe as the bursts become fainter. This effect is likely due to the lack of statistics
encountered when the net burst count rate decreases (making detections statistically weaker) as the persistent
flux rises.


\begin{figure}
\begin{center}
\includegraphics[width=8.5cm]{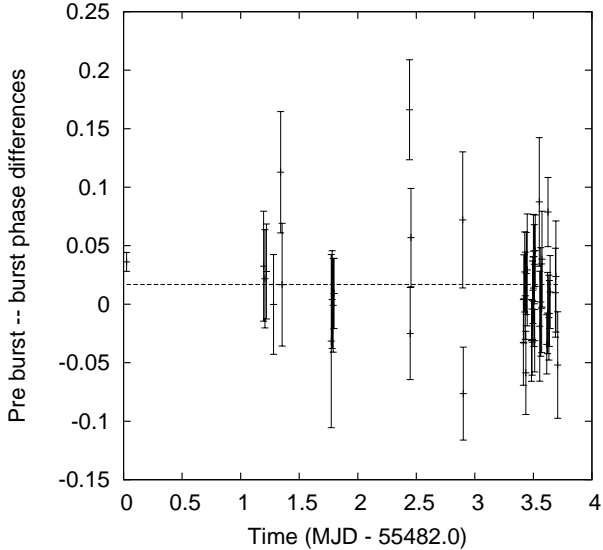}
\caption{Phase difference between the pulse in the persistent emission 
and during the burst period.}\label{fig:phase}
\end{center}
\end{figure}

In order to ascertain the phase relation between the coherent
  pulsations observed during the {\it persistent} emission and burst
  oscillations, we reconstructed pulse profiles over 100 s long
  intervals before each of the bursts, and over 20s intervals after
  the bursts' onset.  We performed an epoch folding around the best estimate of the spin period (see Papitto et al. 2010) for the two
intervals, obtaining two pulse profiles that were fitted according to
a standard harmonic decomposition.  We then obtained the phase
difference between the two profiles and studied the variations of the
phase difference for each burst as a function of time. Such difference
is in general consistent with zero for all examined bursts
(Fig. \ref{fig:phase}). Considering the uncertainties on the
  phase estimates we thus conclude that the bursts oscillations
  originate from a region which is very close to the NS polar caps.

\section{Discussion and conclusions}\label{Sec:disc}

In this paper we report on the bursting behavior of the newly 
discovered X-ray pulsar IGR J17480-2446.

The  observed spin  period and  the magnetic  field estimated  for IGR
J17480-2446 (see \citealt{Papitto2011}) place this source in between the
population  of classical  (B  $\gtrsim$  10$^{11}$G, P  $\ge$  0.1 s)  and
millisecond accreting X-ray pulsars (B  = 10$^8$- 10$^9$ G, P = 1.5-10
ms).   This  source   constitutes  therefore  a  \textit{bona  fide}
candidate as link between these two groups,  being a
slow, probably mildly  recycled,  pulsar.  IGR J17480-2446  is  presently  the
bursting source with the longest spin period observed.

\begin{figure}
\begin{center}
\includegraphics[width=8.5cm]{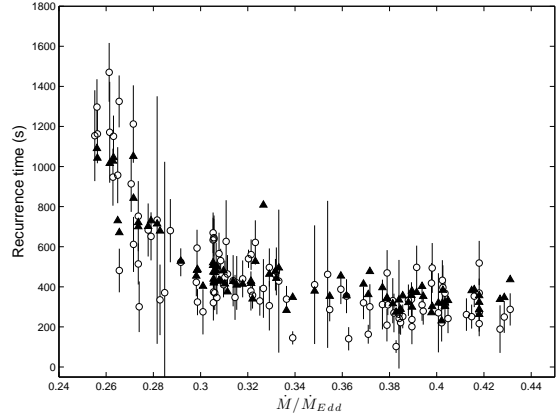}
\caption{Observed (triangles) and predicted (circles) recurrence times  plotted in the function of the measured local accretion rate. Error bars related to the accretion rate are not reported for clarity.}\label{fig:predict}
\end{center}
\end{figure}

To convert flux into luminosity and fluence into energy, we assumed
geometric isotropy of the emission and a source distance of 5.9
kpc. We further express this quantity in units of Eddington luminosity
(=1.8 $\times$10$^{38}$ erg s$^{-1}$ for a 1.4 M$_{\odot}$ NS). From
MJD 55482.00872 to MJD 55486.56326 the source {\it persistent}
luminosity rises from $\sim$2.2$\times$ $d_{5.9}^2$ 10$^{37}$ erg
s$^{-1}$ to $\sim$7.5$\times$10$^{37}$ $d_{5.9}^2$ erg s$^{-1}$.  From
these values, we infer an increase of the mass accretion rate\footnote{The expected values of $\dot{M}/\dot{M}_{Edd}$ are calculated as $\frac{\dot{M}}{\dot{M}_{Edd}} = \frac{\dot{L}}{\dot{L}_{Edd}}$} from
0.1$\dot{M}_{Edd}$ to 0.37$\dot{M}_{Edd}$ (see Fig. \ref{fig:predict}). Consistently with this
behavior the bursts become more frequent and fainter.
\\

In this paper we report on the analysis of 107 X-ray
bursts shown by IGR J17480-2446 during the first four days of the RXTE
monitoring. Since the {\it persistent} bolometric X-ray luminosity
rises by a factor $\sim$4 during this interval we were able to
study the properties of bursts at different mass accretion rates.

The observed bursts have different profiles in the light curve. The rise time varies between
few and $\sim$20 seconds, while the decay time ($\tau$)
spans the range between $\sim$20 and $\sim$100 seconds. The firsts, more intense bursts (MJD 55483 and 55484), show a typical Type-I X-Ray burst profile, featuring a linear rise and an exponential decay following the burst peak. The subsequent fainter bursts present a less clear profile, with rise times in some cases similar to the decay times. For some of the fainter bursts a clear modeling of the profile was difficult because the excursion in intensity above the continuum due to the burst occurrence was comparable with the persistent emission variations.

The pulsation does not show any drifting of the
frequency during the decay phase of the bursts. This phenomenon is
also observed in bursting millisecond pulsars (XTE J1814– 338, \citealt{MS03}; HETE J1900.1–2455,
\citealt{Kaaret2006}; IGR J17511–-3057
\citealt{Altamirano2010b}) and can be explained in terms of a
non-significant expansion of the NS atmosphere during the
thermonuclear combustion. Studying the spectral evolution during the first burst, no photospheric radius expansion is indeed observed (see Fig. \ref{fig:profiles}). 

Burst oscillations are always phase locked within 0.2 phase units, indicating how the ignition begins in a region
  not far from the NS polar caps. The radius inferred from spectral
fitting for the first and most intense burst is 3.2$\pm$0.5 km. Also taking into account the corrective
factors to translate this value into a physical size on the NS surface (see below), the obtained radius is compatible, within the uncertainties, with the NS radius.
\\


The recurrence times of X-ray bursts was observed to decrease from
$\geq$26 min to $\simeq$5 min
in anti-correlation with the persistent X-ray flux. The shortest recurrence time so far observed 
in IGR J17480-2446 is 3.3 min and it is seen in MJD 55487 (Obs. 95437-01-06-00, not considered in this work). The shortest recurrence time previously reported is
3.8 minutes (see \citealt{Keek2010}).

The very short recurrence time observed from the
bursts shown IGR J17480-2446, together with the unusual spectral and
morphological properties, casted doubts on the mechanism powering such
bursts. The absence of a significant cooling from all the bursts
except than for the first brightest one and the unprecedented
observations of long trains of bursts with very short recurrence time
led \cite{Galloway2010} to argue that IGR J17480-2446 could be an
analogue of GRO J1744-28 in showing Type-II bursts powered by
accretion instabilities.

In this paper we have analyzed the bursts shown by IGR J1748-2446 during the first four days of RXTE observations, and we were able to unambiguously measure the recurrence time for a large number
of them. 
We considered the measures of burst fluences and of the
{\it persistent} flux before each burst and we estimated the ratio of the accretion energy liberated into nuclear energy during the bursts,
$\alpha=F_{pers}\:t_{rec}/\mathcal{F}_{burst}$. The average value we have
obtained ($\alpha$= 96$\pm$3) strongly favours the hypothesis these bursts are
due to unstable thermonuclear burning of the H/He accreted on the NS
surface layer. As a matter of fact, under this hypothesis and assuming that all the
available H/He nuclear fuel is burnt during bursts such ratio is expected to be 
$\alpha=Q_{grav}/Q_{nuc} (1+z)$, where $Q_{grav}=GM/R_{NS}=180$ MeV is
the energy liberated per accreted nucleon for a 1.4 M$_{\odot}$ NS
with a radius of 10 km, $1+z=(1-2GM/R_{NS}c^2)^{-1/2}$ measures the
gravitational red shift at the NS surface, and $Q_{nuc}=1.6 + 4<X>$
MeV/nucleon is the energy released during thermonuclear burning of a
mixture of H and He. Here, $<X>$ is the mass fraction of hydrogen
which may not be depleted during the stable burning phase in between
bursts and therefore burns with He during the flash (see e.g.
\citealt{Galloway2008}). The average estimate of $\alpha=96\pm3$ we have given for
the bursts shown by {\fullsrc} is compatible with the burst being of Type-I and indicates 
$<X>$=0.22(2) for (1 + z) = 1.31. In addition, despite the lower statistic of the bursts observed at higher accretion rates, the values of alpha coming from the fainter bursts are consistent with the ones found for lower accretion rate regimes (see Fig. \ref{fig:alfa_grap}), even though the statistics is not enough to clearly identify fluctuations of the values of $\alpha$ driven by the variations of $\dot{m}$. However, the value of $\alpha$ is always close to 96, still indicating the thermonuclear origin as the most probable for all the bursts. 
It is worth to note that Type-II X-Ray bursters such as MBX 1730-335
and GRO J1744-28 show values of alpha of the order unity ($\leq$1.4, \citealt{Kunieda1984},
and $\leq$4, \citealt{Lewin1976} respectively). Similar reasoning led Chakraborty
\& Bhatthacharya (2010) and \cite{Chakraborty2011} to question the interpretation of these bursts
in terms of accretion instabilities\footnote{Type-II bursts observed from MXB 1730-335 (the Rapid Burster) and the accreting X-ray pulsar GRO J1744-28 \citep{Kouveliotou1996} are thought to be produced by spasmodic episodes of accretion, probably triggered by thermal instability at the inner edge of the accretion disk.}. 
 We thus conclude that the most probable interpretation for such bursts is in
terms of helium ignition in a layer still partly composed by H, as
the mass accretion rate is too large to deplete it completely during
the phase of {\it persistent} emission. This conclusion is furthermore supported by the fact that, as stated by \cite{Cumming2000}, for accretion rates $\geq$ 2$\times$ 10$^{-10}$ M$_{sun}$/yr (see \citealt{Bildsten1998} and references therein), the accumulating hydrogen is thermally stable and burns via the hot CNO cycle of \cite{Hoyle1965}.

Such a conclusion about the composition of the burst material also fits well
into theoretical expectations for local accretion rates
$0.1\dot{m}_{Edd}\simlt\dot{m}\simlt\dot{m}_{Edd}$ (e.g. Fujimoto et
al. 1981). The local accretion rate (defined as the total accretion rate divided by the total surface of the NS) onto the NS in {\fullsrc} can be
in fact estimated from the persistent X-ray luminosity as $\dot{m}
\simeq 0.24
(A_{burst}/A_{NS})d_{5.9}^2 \dot{m}_{Edd}$, where $A_{NS}$ and
$A_{burst}$ are the area of the NS surface and that of the region
where the burst is ignited, respectively.  Further evaluating the ignition depth from the burst energetics,
y$_{ign}=\mathcal{F}_{burst}(1+z) (d/r)^2 Q_{nuc}^{-1}$, one obtains that the expected
recurrence time, $\Delta$t=$(y_{ign}/<\dot{m}>) (1+z)$, fits well the observed
values for Q$_{nuc}$=2.5 Mev/nucl. The values that we found are reported in
Tab. \ref{tab:burst} and plotted together with the observed recurrence times in
Fig. \ref{fig:predict}.

Based on the results presented above, combined with the considerations we made according to \cite{Sztajno1986} (see Sec. \ref{Sec:Obs}), we believe that the bursts from IGR J17480-2446, despite their unusual properties, are type-I bursts and that the ambiguity in their classification possibly arises from the presence of a blackbody component in the persistent flux and the relatively low statistics of the spectrum during the bursts following the first one (see. \citealt{VP1986}). 
\\

None of the bursts of {\fullsrc} shows photospheric radius expansion. The accurate knowledge of the distance to Terzan 5
  makes the estimate of the blackbody radius observed during bursts
  very appealing (see e.g. \citealt{Ozel2009}). A detailed spectral
analysis of the brightest burst indicates a blackbody radius of
$R_{app}^{\infty}=(3.3\pm0.5)$ d$_{5.9}$ km. Such a value can be
translated in an effective radius taking into account the spectral
hardening induced by Compton scattering in the NS atmosphere and the
gravitational redshift, $R=R_{app}^{\infty}f_C^2(1+z)^{-1}$. For a
color temperature of the order of 2 keV like the one observed, the
hardening factor $f_C$ was estimated to lie in between 1.33 and 1.84
by \cite{Madej2004}, with the larger values appropriate for NS with
a smaller gravitational acceleration at the surface. Further taking
$1+z=(1-2GM/R_{NS}c^2)^{-1/2}=1.31$ for 1.4 M$_{\odot}$ and 10 km radius NS we obtain an estimate of $R$ in between
$\sim4.5$ and $\sim 9$ km.  The larger value is obtained taking
  a large hardening correction factor which is predicted to be valid
  for a relatively large NS. Also an uncertainty on the distance of the source equal to 0.5 kpc was taken into account. The resulting measure indicates a radius compatible with the typical values for a NS. 
We also measured (see Sec. \ref{Sec:Obs}) the mean apparent radii for selected bursts following the first brightest one (see Tab. \ref{tab:burst}), applying the same corrections that we adopted for the latter. As one can see, also the measures of the radii coming from the fainter bursts confirms that the emitting region during the thermonuclear burst is consistent with the entire NS surface. In addition, this result is in agreement with the fact that the bursts are phase-locked and that there is no variation in the fractional rms associated to the pulse frequency before, during and after the bursts.
\\

Being the NS in {\fullsrc} a pulsar, it is worth to ascertain if the magnetic field is able to confine the accreted matter near the magnetic poles. \cite{Papitto2011} estimated an upper limit on the magnetic field of $\sim 2.4\times 10^{10}$ G considering the lowest flux at which pulsations have been observed. Following the model described in \cite{Brown1998} it results that a similar field can confine the accreted matter near a 5 km magnetic cap up to a column density of $\simeq 3\times 10^8$ g cm$^{-2}$ (see \citealt{Cumming2000}). Such a radius is consistent with the vales coming from the spectral analysis (see Sec. \ref{Sec:Obs}). Considering the relation given above, the column depth at which the bursts of {\fullsrc} ignite can be estimated to lie within 0.05$\times 10^{8}$d$_{5.9}^2$ and 0.5 $\times 10^{8}$ d$_{5.9}^2$g cm$^{(-2)}$, where we have used  $Q_{nuc}$ = 2.5 Mev /nucleon. It is thus possible that, if the magnetic field is in excess of $10^{10}$ G, the radius observed during the type I X-ray bursts reflects only a fraction of the NS surface around the polar caps rather than the entire NS radius. Assuming, that the bursts are ignited onto a smaller fraction of the NS surface (A$_{burst}$) the estimate of the ignition depth grows by a factor ($A_{burst}/A_{NS})^{-2}$, so that values closer to the critical threshold at which helium is thought to start burning unstably on the NS surface, $\geq 6.8 \times 10^8$  (see e.g. \citealt{Cumming2000}) are obtained as soon as ($A_{burst}/A_{NS}$) $\leq$0.5. A confinement of the burst ignition region would straightforwardly help in explaining the very short recurrence times observed in between the fainter bursts. However, as noted before, this interpretation would require a magnetic field larger than the one inferred from the observations. For this reason the data coming from IGR J17480-2446 point out the existence of an interpretation problem that cannot be completely solved by the currently accepted model describing accretion and thermonuclear production mechanism.



In a few LMXBs, oscillations with a period of 100-150 seconds (mHz
QPO) were discovered and associated to marginally stable nuclear
burning on the NS surface (see \citealt{Heger2007};
\citealt{A08}). They are observed only in the luminosity range 0.5-1.5
$\times 10^{37}$ erg s$^{-1}$. The small bursts reported here, with a
recurrence time between 300 and 1000 s and associated to a persistent
luminosity in the range 2.2--7.3 $\times 10^{37}$ erg s$^{-1}$ are
intermediate between full-fledged type-I X-ray bursts and mHz QPOs,
providing an ideal laboratory to study the properties of nuclear
burning on the surface of accreting NSs.

The short recurrence time observed in IGR J17480-2446  is particularly interesting not only because
it is the shortest observed until now. Type-I X-Ray Bursts with very short recurrence times have been studied before (see eg. \citealt{Keek2009}, \citealt{Galloway2004}, \citealt{Boirin2007}). Even though different ideas have been put forward to explain this rare bursting behavior, the short recurrence time still remains an open issue in the theory of thermonuclear X-Ray bursts. In particular, the extreme behavior of IGR J17480-2446 puts the source in accretion regimes well beyond the ones currently investigated by the theory of thermonuclear X-Ray Bursts (see \citealt{Narayan2003} for a detailed study on thermonuclear stability of the accreted matter onto NSs). It is worth to notice that before the discovery of IGR J17480-2446, a maximum of 4 consecutive bursts was observed (see \citealt{Keek2010}). IGR J17480-2446 have shown several tens of Type-I X-Ray bursts with the shortest recurrence time ever observed.
It has been argued that short recurrence time could be due to multi-dimensional effects, such as the confinement of accreted material on different parts of the surface, possibly
as the result of a magnetic field (e.g., \citealt{Melatos2005}; \citealt{Lamb2009}). However, this scenario seems to be ruled out for the case of IGR J17480-2446 by the radius measures of the emitting region, which suggest that a region comparable with the whole NS surface is burning during the bursts.  In addition, the magnetic field of the source appears to be too weak (upper limit $\sim 2.4\times 10^{10}$) to confine the burning matter in particular regions of the NS surface (see above). 
Furthermore, magnetic confinement of accreted material would not be a plausible explanation for the short recurrence time of 3.8 min observed for 4U 1705-44 (see \citealt{Keek2010}), which
does not show coherent pulsations, but it may show a very soft spectrum,
thus indicating a physically weak interacting magnetic field. Also the idea of a burning layer with an unburned layer on
top has been investigated. According to \cite{Fujimoto1988}, after
the first layer flashes, the second layer could be mixed
down to the depth where a thermonuclear runaway occurs thanks to rotational hydrodynamic
instabilities (\citealt{Fujimoto1988}) or by instabilities due to
a rotationally induced magnetic field (\citealt{Piro2007}; \citealt{Keek2009}). 
Being a pulsar and showing burst oscillations, IGR J17480-2446 belongs to a small group of
sources for which the spin period is known and which show short
recurrence time thermonuclear bursts (see \citealt{Keek2010}, Table
1). All these sources proved to be fast spinning NSs with $\nu_{spin}
\ge$ 500 Hz. IGR J17480-2446 is the first NS showing short recurrence
time bursts {\it and} a low spin frequency (11 Hz). This fact
demonstrates that, contrarily to what has been thought until now, fast
rotation is \emph{not} required for the occurrence of multiple-burst
events. For this reason, models predicting multiple burst events
produced by rotationally induced mixing due to fast spinning frequency
should be revised accordingly (see eg. \citealt{Fujimoto1988},
\citealt{Spruit2002}, \citealt{Keek2009}). 
Furthermore, the discovery of IGR J17480-2446 and its bursts oscillations demonstrate the fact that a fast spin frequency is also not required for bursts oscillations to be observed.  Since the burst characteristics (profile, rise and decay times) and their energetics do not change even in long spin period regimes, we can conclude that the spin frequency of the NS does not affect or only marginally affects the burst production mechanism.
\\

We conclude that thanks to its rare and unique behavior IGR J17480-2446 constitutes an ideal laboratory to investigate in detail the mechanisms that triggers thermonuclear bursts and to test the validity of the theoretical models describing the occurrence of very short recurrence times Type-I X-Ray bursts.

\section*{acknowledgements}

 \noindent This work is supported by the Italian Space Agency, ASI-
INAF I/088/06/0 contract for High Energy Astrophysics, as well as by the operating program of Regione Sardegna (European Social Fund 2007-2013), L.R.7/2007, Promotion of scientific research and technological innovation in Sardinia). The research leading to these results has received funding from the   
European Community's Seventh Framework Programme (FP7/2007-2013) under   
grant agreement number ITN 215212. \textquotedblleft Black Hole   
Universe\textquotedblright.

\bibliographystyle{mn2e.bst}
\bibliography{1748_Motta.bib} 

\begin{thebibliography}{}

\bibitem[\protect\citeauthoryear{{Altamirano}, {Casella}, {Patruno}, {Wijnands}
  \& {van der Klis}}{{Altamirano} et~al.}{2008}]{A08}
{Altamirano} D.,  {Casella} P.,  {Patruno} A.,  {Wijnands} R.,    {van der
  Klis} M.,  2008, \apjl, 674, L45

\bibitem[\protect\citeauthoryear{{Altamirano} \& {Watts}}{{Altamirano} \&
  {Watts}}{2010}]{Altamirano2010a}
{Altamirano} D.,  {Watts} A.,  2010, The Astronomer's Telegram, 2932, 1

\bibitem[\protect\citeauthoryear{{Altamirano}, {Watts}, {Linares}, {Markwardt},
  {Strohmayer} \& {Patruno}}{{Altamirano} et~al.}{2010a}]{Altamirano2010}
{Altamirano} D.,  {Watts} A.,  {Linares} M.,  {Markwardt} C.~B.,  {Strohmayer}
  T.,    {Patruno} A.,  2010a, \mnras, pp 1363--+

\bibitem[\protect\citeauthoryear{{Altamirano}, {Watts}, {Linares}, {Markwardt},
  {Strohmayer} \& {Patruno}}{{Altamirano} et~al.}{2010b}]{Altamirano2010b}
{Altamirano} D.,  {Watts} A.,  {Linares} M.,  {Markwardt} C.~B.,  {Strohmayer}
  T.,    {Patruno} A.,  2010b, \mnras, pp 1363--+

\bibitem[\protect\citeauthoryear{{Bildsten}}{{Bildsten}}{1998}]{Bildsten1998}
{Bildsten} L.,  1998, \apjl, 501, L89+

\bibitem[\protect\citeauthoryear{{Boirin}, {Keek}, {M{\'e}ndez}, {Cumming},
  {in't Zand}, {Cottam}, {Paerels} \& {Lewin}}{{Boirin}
  et~al.}{2007}]{Boirin2007}
{Boirin} L.,  {Keek} L.,  {M{\'e}ndez} M.,  {Cumming} A.,  {in't Zand}
  J.~J.~M.,  {Cottam} J.,  {Paerels} F.,    {Lewin} W.~H.~G.,  2007, \aap, 465,
  559

\bibitem[\protect\citeauthoryear{{Brown} \& {Bildsten}}{{Brown} \&
  {Bildsten}}{1998}]{Brown1998}
{Brown} E.~F.,  {Bildsten} L.,  1998, \apj, 496, 915

\bibitem[\protect\citeauthoryear{{Chakraborty} \&
  {Bhattacharyya}}{{Chakraborty} \& {Bhattacharyya}}{2011}]{Chakraborty2011}
{Chakraborty} M.,  {Bhattacharyya} S.,  2011, ArXiv e-prints

\bibitem[\protect\citeauthoryear{{Cumming} \& {Bildsten}}{{Cumming} \&
  {Bildsten}}{2000}]{Cumming2000}
{Cumming} A.,  {Bildsten} L.,  2000, \apj, 544, 453

\bibitem[\protect\citeauthoryear{{Fujimoto}}{{Fujimoto}}{1988}]{Fujimoto1988}
{Fujimoto} M.~Y.,  1988, \aap, 198, 163

\bibitem[\protect\citeauthoryear{{Galloway}, {Cumming}, {Kuulkers}, {Bildsten},
  {Chakrabarty} \& {Rothschild}}{{Galloway} et~al.}{2004}]{Galloway2004}
{Galloway} D.~K.,  {Cumming} A.,  {Kuulkers} E.,  {Bildsten} L.,  {Chakrabarty}
  D.,    {Rothschild} R.~E.,  2004, \apj, 601, 466

\bibitem[\protect\citeauthoryear{{Galloway}, {Lin}, {Chakrabarty} \&
  {Hartman}}{{Galloway} et~al.}{2010}]{Galloway2010}
{Galloway} D.~K.,  {Lin} J.,  {Chakrabarty} D.,    {Hartman} J.~M.,  2010,
  \apjl, 711, L148

\bibitem[\protect\citeauthoryear{{Galloway}, {Muno}, {Hartman}, {Psaltis} \&
  {Chakrabarty}}{{Galloway} et~al.}{2008}]{Galloway2008}
{Galloway} D.~K.,  {Muno} M.~P.,  {Hartman} J.~M.,  {Psaltis} D.,
  {Chakrabarty} D.,  2008, \apjs, 179, 360

\bibitem[\protect\citeauthoryear{{Heger}, {Cumming} \& {Woosley}}{{Heger}
  et~al.}{2007}]{Heger2007}
{Heger} A.,  {Cumming} A.,    {Woosley} S.~E.,  2007, \apj, 665, 1311

\bibitem[\protect\citeauthoryear{{Hoyle} \& {Fowler}}{{Hoyle} \&
  {Fowler}}{1965}]{Hoyle1965}
{Hoyle} F.,  {Fowler} W.~A.,  1965, pp 17--+

\bibitem[\protect\citeauthoryear{{Kaaret}, {Morgan}, {Vanderspek} \&
  {Tomsick}}{{Kaaret} et~al.}{2006}]{Kaaret2006}
{Kaaret} P.,  {Morgan} E.~H.,  {Vanderspek} R.,    {Tomsick} J.~A.,  2006,
  \apj, 638, 963

\bibitem[\protect\citeauthoryear{{Keek}, {Galloway}, {in't Zand} \&
  {Heger}}{{Keek} et~al.}{2010}]{Keek2010}
{Keek} L.,  {Galloway} D.~K.,  {in't Zand} J.~J.~M.,    {Heger} A.,  2010,
  \apj, 718, 292

\bibitem[\protect\citeauthoryear{{Keek}, {Langer} \& {in't Zand}}{{Keek}
  et~al.}{2009}]{Keek2009}
{Keek} L.,  {Langer} N.,    {in't Zand} J.~J.~M.,  2009, \aap, 502, 871

\bibitem[\protect\citeauthoryear{{Kouveliotou}, {van Paradijs}, {Fishman},
  {Briggs}, {Kommers}, {Harmon}, {Meegan} \& {Lewin}}{{Kouveliotou}
  et~al.}{1996}]{Kouveliotou1996}
{Kouveliotou} C.,  {van Paradijs} J.,  {Fishman} G.~J.,  {Briggs} M.~S.,
  {Kommers} J.,  {Harmon} B.~A.,  {Meegan} C.~A.,    {Lewin} W.~H.~G.,  1996,
  \nat, 379, 799

\bibitem[\protect\citeauthoryear{{Kunieda}, {Tawara}, {Hayakawa}, {Nagase},
  {Inoue}, {Kawai}, {Makino}, {Makishima}, {Matsuoka}, {Murakami}, {Oda},
  {Ogawara}, {Ohashi} \& {Waki}}{{Kunieda} et~al.}{1984}]{Kunieda1984}
{Kunieda} H.,  {Tawara} Y.,  {Hayakawa} S.,  {Nagase} F.,  {Inoue} H.,  {Kawai}
  N.,  {Makino} F.,  {Makishima} K.,  {Matsuoka} M.,  {Murakami} T.,  {Oda} M.,
   {Ogawara} Y.,  {Ohashi} T.,    {Waki} I.,  1984, \pasj, 36, 807

\bibitem[\protect\citeauthoryear{{Lamb}, {Boutloukos}, {Van Wassenhove},
  {Chamberlain}, {Lo}, {Clare}, {Yu} \& {Miller}}{{Lamb}
  et~al.}{2009}]{Lamb2009}
{Lamb} F.~K.,  {Boutloukos} S.,  {Van Wassenhove} S.,  {Chamberlain} R.~T.,
  {Lo} K.~H.,  {Clare} A.,  {Yu} W.,    {Miller} M.~C.,  2009, \apj, 706, 417

\bibitem[\protect\citeauthoryear{{Lanzoni}, {Ferraro}, {Dalessandro},
  {Mucciarelli}, {Beccari}, {Miocchi}, {Bellazzini}, {Rich}, {Origlia},
  {Valenti}, {Rood} \& {Ransom}}{{Lanzoni} et~al.}{2010}]{Lanzoni2010}
{Lanzoni} B.,  {Ferraro} F.~R.,  {Dalessandro} E.,  {Mucciarelli} A.,
  {Beccari} G.,  {Miocchi} P.,  {Bellazzini} M.,  {Rich} R.~M.,  {Origlia} L.,
  {Valenti} E.,  {Rood} R.~T.,    {Ransom} S.~M.,  2010, \apj, 717, 653

\bibitem[\protect\citeauthoryear{{Leahy}, {Elsner} \& {Weisskopf}}{{Leahy}
  et~al.}{1983}]{Leahy1983}
{Leahy} D.~A.,  {Elsner} R.~F.,    {Weisskopf} M.~C.,  1983, \apj, 272, 256

\bibitem[\protect\citeauthoryear{{Lewin}, {Doty}, {Clark}, {Rappaport},
  {Bradt}, {Doxsey}, {Hearn}, {Hoffman}, {Jernigan}, {Li}, {Mayer},
  {McClintock}, {Primini} \& {Richardson}}{{Lewin} et~al.}{1976}]{Lewin1976}
{Lewin} W.~H.~G.,  {Doty} J.,  {Clark} G.~W.,  {Rappaport} S.~A.,  {Bradt}
  H.~V.~D.,  {Doxsey} R.,  {Hearn} D.~R.,  {Hoffman} J.~A.,  {Jernigan} J.~G.,
  {Li} F.~K.,  {Mayer} W.,  {McClintock} J.,  {Primini} F.,    {Richardson} J.,
   1976, \apjl, 207, L95

\bibitem[\protect\citeauthoryear{{Madej}, {Joss} \&
  {R{\'o}{\.z}a{\'n}ska}}{{Madej} et~al.}{2004}]{Madej2004}
{Madej} J.,  {Joss} P.~C.,    {R{\'o}{\.z}a{\'n}ska} A.,  2004, \apj, 602, 904

\bibitem[\protect\citeauthoryear{{Markwardt} \& {Swank}}{{Markwardt} \&
  {Swank}}{2003}]{MS03}
{Markwardt} C.~B.,  {Swank} J.~H.,  2003, \iaucirc, 8144, 1

\bibitem[\protect\citeauthoryear{{Melatos} \& {Payne}}{{Melatos} \&
  {Payne}}{2005}]{Melatos2005}
{Melatos} A.,  {Payne} D.~J.~B.,  2005, \apj, 623, 1044

\bibitem[\protect\citeauthoryear{{Muno}, {Chakrabarty}, {Galloway} \&
  {Psaltis}}{{Muno} et~al.}{2002}]{Muno2002}
{Muno} M.~P.,  {Chakrabarty} D.,  {Galloway} D.~K.,    {Psaltis} D.,  2002,
  \apj, 580, 1048

\bibitem[\protect\citeauthoryear{{Narayan} \& {Heyl}}{{Narayan} \&
  {Heyl}}{2003}]{Narayan2003}
{Narayan} R.,  {Heyl} J.~S.,  2003, \apj, 599, 419

\bibitem[\protect\citeauthoryear{{{\"O}zel}, {G{\"u}ver} \&
  {Psaltis}}{{{\"O}zel} et~al.}{2009}]{Ozel2009}
{{\"O}zel} F.,  {G{\"u}ver} T.,    {Psaltis} D.,  2009, \apj, 693, 1775

\bibitem[\protect\citeauthoryear{{Papitto}, {D'A{\`i}}, {Motta}, {Riggio},
  {Burderi}, {di Salvo}, {Belloni} \& {Iaria}}{{Papitto}
  et~al.}{2011}]{Papitto2011}
{Papitto} A.,  {D'A{\`i}} A.,  {Motta} S.,  {Riggio} A.,  {Burderi} L.,  {di
  Salvo} T.,  {Belloni} T.,    {Iaria} R.,  2011, \aap, 526, L3+

\bibitem[\protect\citeauthoryear{{Piro} \& {Bildsten}}{{Piro} \&
  {Bildsten}}{2007}]{Piro2007}
{Piro} A.~L.,  {Bildsten} L.,  2007, \apj, 663, 1252

\bibitem[\protect\citeauthoryear{{Pooley}, {Homan} \& {Heinke}}{{Pooley}
  et~al.}{2010}]{Pooley2010}
{Pooley} D.,  {Homan} J.,    {Heinke} C.,  2010, The Astronomer's Telegram,
  2974, 1

\bibitem[\protect\citeauthoryear{{Spruit}}{{Spruit}}{2002}]{Spruit2002}
{Spruit} H.~C.,  2002, \aap, 381, 923

\bibitem[\protect\citeauthoryear{{Strohmayer} \& {Bildsten}}{{Strohmayer} \&
  {Bildsten}}{2006}]{Strohmayer2006}
{Strohmayer} T.,  {Bildsten} L.,  2006, {New views of thermonuclear bursts}

\bibitem[\protect\citeauthoryear{{Strohmayer} \& {Markwardt}}{{Strohmayer} \&
  {Markwardt}}{2010}]{Strohmayer2010}
{Strohmayer} T.~E.,  {Markwardt} C.~B.,  2010, The Astronomer's Telegram, 2929,
  1

\bibitem[\protect\citeauthoryear{{Strohmayer}, {Markwardt} \&
  {Pereira}}{{Strohmayer} et~al.}{2010}]{Strohmayer2010a}
{Strohmayer} T.~E.,  {Markwardt} C.~B.,    {Pereira} D.,  2010, The
  Astronomer's Telegram, 2946, 1

\bibitem[\protect\citeauthoryear{{Strohmayer}, {Markwardt}, {Swank} \& {in't
  Zand}}{{Strohmayer} et~al.}{2003}]{Strohmayer2003}
{Strohmayer} T.~E.,  {Markwardt} C.~B.,  {Swank} J.~H.,    {in't Zand} J.,
  2003, \apjl, 596, L67

\bibitem[\protect\citeauthoryear{{Sztajno}, {van Paradijs}, {Lewin},
  {Langmeier}, {Trumper} \& {Pietsch}}{{Sztajno} et~al.}{1986}]{Sztajno1986}
{Sztajno} M.,  {van Paradijs} J.,  {Lewin} W.~H.~G.,  {Langmeier} A.,
  {Trumper} J.,    {Pietsch} W.,  1986, \mnras, 222, 499

\bibitem[\protect\citeauthoryear{{van Paradijs} \& {Lewin}}{{van Paradijs} \&
  {Lewin}}{1986}]{VP1986}
{van Paradijs} J.,  {Lewin} H.~G.,  1986, \aap, 157, L10

\bibitem[\protect\citeauthoryear{{Woosley}, {Heger}, {Cumming}, {Hoffman},
  {Pruet}, {Rauscher}, {Fisker}, {Schatz}, {Brown} \& {Wiescher}}{{Woosley}
  et~al.}{2004}]{Woosley2004}
{Woosley} S.~E.,  {Heger} A.,  {Cumming} A.,  {Hoffman} R.~D.,  {Pruet} J.,
  {Rauscher} T.,  {Fisker} J.~L.,  {Schatz} H.,  {Brown} B.~A.,    {Wiescher}
  M.,  2004, \apjs, 151, 75

\end{thebibliography}


\onecolumn	
																										
\begin{center}	
\begin{landscape}	
\begin{longtable}{|c|c|c|c|c|c|c|c|c|c|c|}																											
\caption[]{Properties	of	the	107	X-ray	bursts	from	IGR	J17480-2446.																			
Columns	are:	burst number,	start time	(MJD, referred to the Solar System center of mass),	recurrence time	(s),	rise	time	(s),	decay	time	$\tau$	(s),	persistent flux,	burst fluence, $\alpha$, predicted recurrence time (assuming the energy conversion efficiency equal to $\sim$ 2Mev/Nucleon).}\label{tab:burst}\\																									
\endfirsthead																											
\multicolumn{11}{c}%
{{\tablename\	\thetable{}	--	continued	from	previous	page}}	\\																				

\hline																											
																											
No.	&	Obs.ID	&	T$_{start}$	&	$T_{rec}$			&	t$_{rise}$			&	$\tau$			&	Persistent flux			&	Burst Fluence			&	$\alpha$	& $T_{rec} ^{pred}$	\\
 &	&	(MJD)	&	(s)			&	(s)			&	(s)			&	(10$^{-8}$erg cm$^{-2} s^{-1}$)			&	(10$^{-8}$erg cm$^{-2}$)			&				& (s)\\
\hline																											
\hline																											
\endhead																											
\hline	\multicolumn{11}{c}{{Continued	on	next	page}}			\\																				
\endfoot																											
\hline																											
\endlastfoot																											
\hline	
																										
No.	&	Obs.ID	&	T$_{start}$	&	$T_{rec}$			&	t$_{rise}$			&	$\tau$			&	Persistent flux			&	Burst Fluence			&	$\alpha$	& $T_{rec} ^{pred}$	\\
 &	&	(MJD)	&	(s)			&	(s)			&	(s)			&	(10$^{-8}$erg cm$^{-2} s^{-1}$)			&	(10$^{-8}$erg cm$^{-2}$)			&				& (s)\\
\hline																											
\hline

1	&	95437-01-01-00	&	55482.03	&	***			&	5.0	$\pm$	0.1	&	34	$\pm$	1	&	0.52	$\pm$	0.01	&	31.5	$\pm$	0.6	&	*		*	&	6448	$\pm$	167		\\
																																		
2	&	95437-01-02-00	&	55483.20	&	***			&	7	$\pm$	1	&	23	$\pm$	3	&	1.13	$\pm$	0.08	&	15.7	$\pm$	1.1	&	*		*	&	1470	$\pm$	147		\\
3	&	95437-01-02-00	&	55483.21	&	1052	$\pm$	1	&	11	$\pm$	1	&	17	$\pm$	3	&	1.17	$\pm$	0.13	&	13.4	$\pm$	1.5	&	92	$\pm$	10	&	1212	$\pm$	193		\\
4	&	95437-01-02-00	&	55483.22	&	1016	$\pm$	2	&	11	$\pm$	2	&	20	$\pm$	2	&	1.13	$\pm$	0.17	&	12.5	$\pm$	1.9	&	92	$\pm$	14	&	1171	$\pm$	252		\\
5	&	95437-01-02-00	&	55483.26	&	***			&	9	$\pm$	1	&	21	$\pm$	3	&	1.10	$\pm$	0.15	&	12.0	$\pm$	1.7	&	*		*	&	1154	$\pm$	227		\\
6	&	95437-01-02-00	&	55483.27	&	1043	$\pm$	1	&	12	$\pm$	1	&	16	$\pm$	3	&	1.11	$\pm$	0.10	&	12.2	$\pm$	1.1	&	95	$\pm$	8	&	1163	$\pm$	147		\\
7	&	95437-01-02-00	&	55483.28	&	1090	$\pm$	1	&	8	$\pm$	1	&	22	$\pm$	3	&	1.11	$\pm$	0.08	&	13.6	$\pm$	1.0	&	89	$\pm$	7	&	1297	$\pm$	139		\\
8	&	95437-01-02-00	&	55483.33	&	***			&	12	$\pm$	1	&	25	$\pm$	3	&	1.15	$\pm$	0.08	&	14.4	$\pm$	1.0	&	*		*	&	1325	$\pm$	129		\\
9	&	95437-01-02-00	&	55483.34	&	1028	$\pm$	1	&	10	$\pm$	1	&	17	$\pm$	2	&	1.14	$\pm$	0.12	&	10.2	$\pm$	1.1	&	115	$\pm$	12	&	947	$\pm$	142		\\
10	&	95437-01-02-00	&	55483.35	&	1045	$\pm$	2	&	11	$\pm$	2	&	17	$\pm$	2	&	1.14	$\pm$	0.07	&	12.4	$\pm$	0.8	&	96	$\pm$	6	&	1151	$\pm$	103		\\
																																		
11	&	95437-01-02-01	&	55483.72	&	***			&	17	$\pm$	2	&	16	$\pm$	2	&	1.17	$\pm$	0.13	&	10.1	$\pm$	1.1	&	*		*	&	914	$\pm$	140		\\
12	&	95437-01-02-01	&	55483.73	&	730	$\pm$	3	&	14	$\pm$	2	&	21	$\pm$	3	&	1.15	$\pm$	0.12	&	10.4	$\pm$	1.1	&	81	$\pm$	8	&	957	$\pm$	140		\\
13	&	95437-01-02-01	&	55483.74	&	842	$\pm$	3	&	13	$\pm$	2	&	20	$\pm$	2	&	1.17	$\pm$	0.19	&	6.8	$\pm$	1.1	&	146	$\pm$	23	&	612	$\pm$	137		\\
14	&	95437-01-02-01	&	55483.75	&	721	$\pm$	2	&	12	$\pm$	2	&	21	$\pm$	3	&	1.18	$\pm$	0.17	&	5.7	$\pm$	0.8	&	148	$\pm$	21	&	514	$\pm$	103		\\
15	&	95437-01-02-01	&	55483.78	&	***			&	14	$\pm$	2	&	15	$\pm$	3	&	1.23	$\pm$	1.53	&	4.3	$\pm$	5.4	&	*		*	&	371	$\pm$	653		\\
16	&	95437-01-02-01	&	55483.79	&	700	$\pm$	2	&	14	$\pm$	1	&	14	$\pm$	2	&	1.18	$\pm$	0.19	&	8.4	$\pm$	1.4	&	99	$\pm$	16	&	753	$\pm$	175		\\
17	&	95437-01-02-01	&	55483.80	&	702	$\pm$	2	&	9	$\pm$	1	&	18	$\pm$	2	&	1.20	$\pm$	0.16	&	7.8	$\pm$	1.1	&	109	$\pm$	15	&	684	$\pm$	132		\\
18	&	95437-01-02-01	&	55483.81	&	679	$\pm$	2	&	16	$\pm$	2	&	13	$\pm$	2	&	1.22	$\pm$	0.45	&	3.9	$\pm$	1.4	&	215	$\pm$	80	&	335	$\pm$	176		\\
19	&	95437-01-02-01	&	55483.85	&	***			&	15	$\pm$	2	&	10	$\pm$	2	&	1.18	$\pm$	0.35	&	3.3	$\pm$	1.0	&	*		*	&	300	$\pm$	126		\\
20	&	95437-01-02-01	&	55483.86	&	671	$\pm$	3	&	9	$\pm$	2	&	21	$\pm$	3	&	1.15	$\pm$	0.18	&	5.2	$\pm$	0.8	&	147	$\pm$	24	&	481	$\pm$	109		\\
21	&	95437-01-02-01	&	55483.86	&	730	$\pm$	3	&	12	$\pm$	2	&	19	$\pm$	3	&	1.21	$\pm$	0.16	&	7.5	$\pm$	1.0	&	119	$\pm$	15	&	652	$\pm$	120		\\
22	&	95437-01-02-01	&	55483.87	&	714	$\pm$	2	&	15	$\pm$	1	&	17	$\pm$	2	&	1.22	$\pm$	0.73	&	8.4	$\pm$	5.0	&	103	$\pm$	61	&	733	$\pm$	618		\\
																																		
23	&	95437-01-03-00	&	55484.44	&	***			&	4	$\pm$	0	&	12	$\pm$	2	&	1.24	$\pm$	0.20	&	8.0	$\pm$	1.3	&	*		*	&	680	$\pm$	158		\\
24	&	95437-01-03-00	&	55484.44	&	501	$\pm$	2	&	12	$\pm$	2	&	13	$\pm$	3	&	1.33	$\pm$	0.23	&	4.4	$\pm$	0.7	&	153	$\pm$	26	&	346	$\pm$	83		\\
25	&	95437-01-03-00	&	55484.45	&	521	$\pm$	4	&	15	$\pm$	3	&	16	$\pm$	3	&	1.32	$\pm$	0.16	&	7.9	$\pm$	1.0	&	87	$\pm$	11	&	633	$\pm$	111		\\
26	&	95437-01-03-00	&	55484.46	&	529	$\pm$	5	&	11	$\pm$	4	&	17	$\pm$	3	&	1.26	$\pm$	0.12	&	6.2	$\pm$	0.6	&	107	$\pm$	10	&	522	$\pm$	70		\\
27	&	95437-01-03-00	&	55484.46	&	487	$\pm$	6	&	17	$\pm$	5	&	14	$\pm$	4	&	1.29	$\pm$	0.18	&	4.0	$\pm$	0.6	&	159	$\pm$	23	&	324	$\pm$	66		\\
																																		
28	&	95437-01-03-01	&	55484.57	&	***			&	12	$\pm$	3	&	19	$\pm$	4	&	1.36	$\pm$	0.31	&	4.5	$\pm$	1.0	&	*		*	&	347	$\pm$	112		\\
29	&	95437-01-03-01	&	55484.58	&	439	$\pm$	4	&	13	$\pm$	2	&	20	$\pm$	5	&	1.32	$\pm$	0.22	&	4.7	$\pm$	0.8	&	124	$\pm$	21	&	374	$\pm$	88		\\
30	&	95437-01-03-01	&	55484.59	&	482	$\pm$	3	&	9	$\pm$	1	&	25	$\pm$	4	&	1.33	$\pm$	0.25	&	6.2	$\pm$	1.2	&	104	$\pm$	19	&	491	$\pm$	129		\\
31	&	95437-01-03-01	&	55484.59	&	483	$\pm$	3	&	14	$\pm$	2	&	16	$\pm$	3	&	1.29	$\pm$	0.14	&	7.2	$\pm$	0.8	&	86	$\pm$	9	&	593	$\pm$	92		\\
32	&	95437-01-03-01	&	55484.60	&	454	$\pm$	3	&	16	$\pm$	2	&	13	$\pm$	2	&	1.29	$\pm$	0.15	&	5.2	$\pm$	0.6	&	114	$\pm$	13	&	423	$\pm$	70		\\
																																		
33	&	95437-01-03-02	&	55484.63	&	***			&	15	$\pm$	3	&	32	$\pm$	5	&	1.34	$\pm$	0.31	&	7.9	$\pm$	1.8	&	*		*	&	626	$\pm$	206		\\
34	&	95437-01-03-02	&	55484.63	&	472	$\pm$	4	&	14	$\pm$	2	&	14	$\pm$	3	&	1.32	$\pm$	0.22	&	4.6	$\pm$	0.8	&	135	$\pm$	22	&	370	$\pm$	87		\\
35	&	95437-01-03-02	&	55484.64	&	480	$\pm$	3	&	13	$\pm$	2	&	11	$\pm$	3	&	1.34	$\pm$	0.15	&	5.2	$\pm$	0.6	&	124	$\pm$	14	&	411	$\pm$	63		\\
36	&	95437-01-03-02	&	55484.65	&	507	$\pm$	3	&	8	$\pm$	2	&	29	$\pm$	20	&	1.32	$\pm$	0.14	&	8.3	$\pm$	0.9	&	80	$\pm$	9	&	669	$\pm$	101		\\
37	&	95437-01-03-02	&	55484.65	&	474	$\pm$	11	&	15	$\pm$	11	&	18	$\pm$	6	&	1.32	$\pm$	0.26	&	4.0	$\pm$	0.8	&	156	$\pm$	31	&	321	$\pm$	89		\\
38	&	95437-01-03-02	&	55484.66	&	420	$\pm$	14	&	54	$\pm$	9	&	113	$\pm$	32	&	1.32	$\pm$	0.13	&	8.0	$\pm$	0.8	&	69	$\pm$	7	&	645	$\pm$	93		\\
39	&	95437-01-03-02	&	55484.66	&	528	$\pm$	9	&	19	$\pm$	2	&	22	$\pm$	4	&	1.40	$\pm$	0.18	&	8.2	$\pm$	1.0	&	90	$\pm$	11	&	621	$\pm$	110		\\
																																		
40	&	95437-01-03-03	&	55484.76	&	***			&	18	$\pm$	7	&	10	$\pm$	3	&	1.53	$\pm$	0.86	&	6.7	$\pm$	3.7	&	*		*	&	462	$\pm$	367		\\
41	&	95437-01-03-03	&	55484.76	&	379	$\pm$	8	&	91	$\pm$	4	&	92	$\pm$	2	&	1.51	$\pm$	0.77	&	5.9	$\pm$	3.0	&	98	$\pm$	50	&	411	$\pm$	296		\\
42	&	95437-01-03-03	&	55484.77	&	495	$\pm$	5	&	13	$\pm$	2	&	7	$\pm$	2	&	1.44	$\pm$	0.85	&	5.8	$\pm$	3.4	&	122	$\pm$	72	&	429	$\pm$	357		\\
43	&	95437-01-03-03	&	55484.77	&	463	$\pm$	2	&	3	$\pm$	1	&	41	$\pm$	5	&	1.42	$\pm$	0.19	&	6.7	$\pm$	0.9	&	99	$\pm$	13	&	496	$\pm$	95		\\
44	&	95437-01-03-03	&	55484.78	&	418	$\pm$	5	&	25	$\pm$	5	&	74	$\pm$	18	&	1.39	$\pm$	0.13	&	7.4	$\pm$	0.7	&	79	$\pm$	8	&	560	$\pm$	76		\\
45	&	95437-01-03-03	&	55484.78	&	477	$\pm$	6	&	10	$\pm$	3	&	19	$\pm$	5	&	1.61	$\pm$	0.43	&	4.6	$\pm$	1.2	&	168	$\pm$	45	&	300	$\pm$	113		\\
46	&	95437-01-03-03	&	55484.79	&	454	$\pm$	4	&	20	$\pm$	3	&	11	$\pm$	3	&	1.55	$\pm$	0.21	&	5.7	$\pm$	0.8	&	124	$\pm$	17	&	388	$\pm$	74		\\
47	&	95437-01-03-03	&	55484.79	&	478	$\pm$	4	&	13	$\pm$	3	&	19	$\pm$	4	&	1.43	$\pm$	0.17	&	6.2	$\pm$	0.8	&	110	$\pm$	13	&	460	$\pm$	79		\\
48	&	95437-01-03-03	&	55484.82	&	***			&	12	$\pm$	4	&	20	$\pm$	5	&	1.42	$\pm$	0.41	&	4.1	$\pm$	1.2	&	*		*	&	306	$\pm$	123		\\
49	&	95437-01-03-03	&	55484.83	&	353	$\pm$	11	&	51	$\pm$	10	&	12	$\pm$	4	&	1.53	$\pm$	0.22	&	4.1	$\pm$	0.6	&	131	$\pm$	19	&	286	$\pm$	59		\\
50	&	95437-01-03-03	&	55484.83	&	427	$\pm$	11	&	39	$\pm$	4	&	46	$\pm$	9	&	1.32	$\pm$	0.17	&	8.0	$\pm$	1.0	&	71	$\pm$	9	&	638	$\pm$	118		\\
51	&	95437-01-03-03	&	55484.84	&	415	$\pm$	6	&	23	$\pm$	3	&	14	$\pm$	4	&	1.34	$\pm$	0.18	&	5.7	$\pm$	0.7	&	98	$\pm$	13	&	447	$\pm$	83		\\
52	&	95437-01-03-03	&	55484.84	&	403	$\pm$	4	&	3	$\pm$	1	&	82	$\pm$	17	&	1.30	$\pm$	0.38	&	3.4	$\pm$	1.0	&	155	$\pm$	45	&	275	$\pm$	113		\\
53	&	95437-01-03-03	&	55484.85	&	409	$\pm$	3	&	19	$\pm$	3	&	10	$\pm$	2	&	1.36	$\pm$	0.31	&	5.4	$\pm$	1.2	&	103	$\pm$	23	&	419	$\pm$	134		\\
54	&	95437-01-03-03	&	55484.85	&	432	$\pm$	6	&	24	$\pm$	5	&	37	$\pm$	5	&	1.33	$\pm$	0.17	&	7.1	$\pm$	0.9	&	81	$\pm$	11	&	566	$\pm$	104		\\
55	&	95437-01-03-03	&	55484.86	&	429	$\pm$	6	&	4	$\pm$	3	&	48	$\pm$	13	&	1.33	$\pm$	0.10	&	6.7	$\pm$	0.5	&	86	$\pm$	7	&	531	$\pm$	59		\\
56	&	95437-01-03-03	&	55484.89	&	***			&	4	$\pm$	4	&	34	$\pm$	7	&	1.39	$\pm$	0.08	&	7.1	$\pm$	0.4	&	*		*	&	540	$\pm$	46		\\
57	&	95437-01-03-03	&	55484.90	&	808	$\pm$	5	&	10	$\pm$	3	&	14	$\pm$	4	&	1.41	$\pm$	0.37	&	5.2	$\pm$	1.4	&	218	$\pm$	58	&	392	$\pm$	147		\\
58	&	95437-01-03-03	&	55484.90	&	414	$\pm$	4	&	7	$\pm$	3	&	45	$\pm$	10	&	1.37	$\pm$	0.20	&	5.7	$\pm$	0.8	&	100	$\pm$	15	&	440	$\pm$	92		\\
59	&	95437-01-03-03	&	55484.91	&	428	$\pm$	15	&	54	$\pm$	14	&	58	$\pm$	15	&	1.39	$\pm$	0.25	&	5.0	$\pm$	0.9	&	120	$\pm$	22	&	379	$\pm$	98		\\
60	&	95437-01-03-03	&	55484.91	&	427	$\pm$	15	&	17	$\pm$	4	&	9	$\pm$	4	&	1.36	$\pm$	0.28	&	5.5	$\pm$	1.2	&	105	$\pm$	22	&	430	$\pm$	128		\\
61	&	95437-01-03-03	&	55484.91	&	375	$\pm$	5	&	43	$\pm$	4	&	18	$\pm$	4	&	1.35	$\pm$	0.19	&	5.9	$\pm$	0.9	&	86	$\pm$	12	&	464	$\pm$	94		\\
62	&	95437-01-03-03	&	55484.92	&	443	$\pm$	5	&	10	$\pm$	4	&	111	$\pm$	18	&	1.44	$\pm$	0.25	&	6.5	$\pm$	1.1	&	98	$\pm$	17	&	477	$\pm$	119		\\
																																		
63	&	95437-01-04-00	&	55485.41	&	***			&	27	$\pm$	4	&	9	$\pm$	4	&	1.41	$\pm$	0.22	&	4.4	$\pm$	0.7	&	*		*	&	330	$\pm$	73		\\
64	&	95437-01-04-00	&	55485.42	&	340	$\pm$	4	&	13	$\pm$	2	&	14	$\pm$	3	&	1.39	$\pm$	0.18	&	4.7	$\pm$	0.6	&	100	$\pm$	13	&	358	$\pm$	67		\\
65	&	95437-01-04-00	&	55485.42	&	283	$\pm$	5	&	32	$\pm$	5	&	9	$\pm$	3	&	1.45	$\pm$	0.20	&	4.6	$\pm$	0.6	&	88	$\pm$	12	&	339	$\pm$	66		\\
66	&	95437-01-04-00	&	55485.43	&	356	$\pm$	6	&	40	$\pm$	3	&	146	$\pm$	15	&	1.56	$\pm$	0.26	&	5.2	$\pm$	0.9	&	107	$\pm$	18	&	353	$\pm$	85		\\
67	&	95437-01-04-00	&	55485.43	&	396	$\pm$	4	&	17	$\pm$	3	&	10	$\pm$	3	&	1.63	$\pm$	0.46	&	4.8	$\pm$	1.3	&	134	$\pm$	38	&	312	$\pm$	124		\\
68	&	95437-01-04-00	&	55485.44	&	414	$\pm$	4	&	12	$\pm$	3	&	23	$\pm$	5	&	1.60	$\pm$	0.29	&	4.8	$\pm$	0.9	&	137	$\pm$	25	&	320	$\pm$	82		\\
69	&	95437-01-04-00	&	55485.44	&	362	$\pm$	5	&	15	$\pm$	4	&	20	$\pm$	4	&	1.60	$\pm$	0.33	&	2.5	$\pm$	0.5	&	234	$\pm$	48	&	163	$\pm$	47		\\
70	&	95437-01-04-00	&	55485.44	&	274	$\pm$	7	&	51	$\pm$	6	&	17	$\pm$	4	&	1.66	$\pm$	0.37	&	1.6	$\pm$	0.4	&	285	$\pm$	63	&	102	$\pm$	32		\\
71	&	95437-01-04-00	&	55485.45	&	348	$\pm$	12	&	50	$\pm$	11	&	8	$\pm$	3	&	1.47	$\pm$	0.23	&	2.0	$\pm$	0.3	&	253	$\pm$	42	&	145	$\pm$	33		\\
72	&	95437-01-04-00	&	55485.48	&	***			&	35	$\pm$	4	&	516	$\pm$	99	&	1.57	$\pm$	0.46	&	2.1	$\pm$	0.6	&	*		*	&	141	$\pm$	58		\\
73	&	95437-01-04-00	&	55485.49	&	402	$\pm$	5	&	14	$\pm$	3	&	18	$\pm$	4	&	1.70	$\pm$	0.31	&	5.0	$\pm$	0.9	&	137	$\pm$	25	&	310	$\pm$	80		\\
74	&	95437-01-04-00	&	55485.49	&	342	$\pm$	4	&	18	$\pm$	3	&	30	$\pm$	7	&	1.64	$\pm$	0.28	&	7.3	$\pm$	1.2	&	77	$\pm$	13	&	469	$\pm$	114		\\
75	&	95437-01-04-00	&	55485.50	&	353	$\pm$	4	&	13	$\pm$	3	&	19	$\pm$	4	&	1.70	$\pm$	0.29	&	4.5	$\pm$	0.8	&	134	$\pm$	23	&	279	$\pm$	68		\\
76	&	95437-01-04-00	&	55485.50	&	288	$\pm$	6	&	20	$\pm$	5	&	13	$\pm$	3	&	1.66	$\pm$	0.32	&	3.5	$\pm$	0.7	&	138	$\pm$	27	&	221	$\pm$	60		\\
77	&	95437-01-04-00	&	55485.50	&	335	$\pm$	7	&	34	$\pm$	5	&	16	$\pm$	8	&	1.66	$\pm$	1.21	&	4.1	$\pm$	3.0	&	135	$\pm$	98	&	263	$\pm$	271		\\
78	&	95437-01-04-00	&	55485.51	&	358	$\pm$	6	&	14	$\pm$	4	&	27	$\pm$	9	&	1.68	$\pm$	0.30	&	5.4	$\pm$	1.0	&	112	$\pm$	20	&	338	$\pm$	85		\\
79	&	95437-01-04-00	&	55485.51	&	278	$\pm$	8	&	32	$\pm$	7	&	29	$\pm$	9	&	1.66	$\pm$	0.37	&	3.8	$\pm$	0.8	&	121	$\pm$	27	&	242	$\pm$	76		\\
																																		
80	&	95437-01-04-01	&	55485.55	&	***			&	15	$\pm$	3	&	43	$\pm$	7	&	1.65	$\pm$	0.42	&	4.2	$\pm$	1.1	&	*		*	&	269	$\pm$	96		\\
81	&	95437-01-04-01	&	55485.56	&	356	$\pm$	3	&	17	$\pm$	1	&	111	$\pm$	33	&	1.67	$\pm$	0.28	&	4.0	$\pm$	0.7	&	149	$\pm$	25	&	252	$\pm$	59		\\
82	&	95437-01-04-01	&	55485.56	&	381	$\pm$	3	&	27	$\pm$	3	&	10	$\pm$	5	&	1.79	$\pm$	0.30	&	4.3	$\pm$	0.7	&	161	$\pm$	27	&	251	$\pm$	60		\\
83	&	95437-01-04-01	&	55485.56	&	320	$\pm$	5	&	47	$\pm$	4	&	9	$\pm$	2	&	1.73	$\pm$	0.89	&	4.4	$\pm$	2.3	&	125	$\pm$	65	&	270	$\pm$	197		\\
84	&	95437-01-04-01	&	55485.57	&	345	$\pm$	8	&	24	$\pm$	7	&	42	$\pm$	7	&	1.64	$\pm$	0.45	&	3.2	$\pm$	0.9	&	175	$\pm$	48	&	208	$\pm$	81		\\
85	&	95437-01-04-01	&	55485.57	&	322	$\pm$	8	&	20	$\pm$	5	&	48	$\pm$	17	&	1.68	$\pm$	0.28	&	5.3	$\pm$	0.9	&	102	$\pm$	17	&	333	$\pm$	78		\\
86	&	95437-01-04-01	&	55485.57	&	299	$\pm$	6	&	24	$\pm$	5	&	14	$\pm$	3	&	1.68	$\pm$	0.26	&	3.8	$\pm$	0.6	&	134	$\pm$	21	&	237	$\pm$	51		\\
87	&	95437-01-04-01	&	55485.58	&	316	$\pm$	5	&	18	$\pm$	3	&	33	$\pm$	12	&	1.65	$\pm$	0.40	&	5.2	$\pm$	1.2	&	101	$\pm$	24	&	331	$\pm$	113		\\
88	&	95437-01-04-01	&	55485.61	&	***			&	13	$\pm$	3	&	32	$\pm$	8	&	1.64	$\pm$	0.27	&	4.8	$\pm$	0.6	&	*		*	&	310	$\pm$	52		\\
89	&	95437-01-04-01	&	55485.61	&	287	$\pm$	5	&	39	$\pm$	4	&	32	$\pm$	10	&	1.81	$\pm$	0.37	&	3.7	$\pm$	0.8	&	141	$\pm$	29	&	216	$\pm$	63		\\
90	&	95437-01-04-01	&	55485.61	&	302	$\pm$	7	&	28	$\pm$	6	&	92	$\pm$	24	&	1.72	$\pm$	0.31	&	8.0	$\pm$	1.4	&	65	$\pm$	12	&	494	$\pm$	124		\\
91	&	95437-01-04-01	&	55485.62	&	347	$\pm$	6	&	17	$\pm$	3	&	16	$\pm$	4	&	1.85	$\pm$	0.42	&	4.4	$\pm$	1.0	&	147	$\pm$	33	&	249	$\pm$	79		\\
92	&	95437-01-04-01	&	55485.62	&	321	$\pm$	4	&	22	$\pm$	2	&	19	$\pm$	4	&	1.74	$\pm$	0.19	&	6.0	$\pm$	0.7	&	92	$\pm$	10	&	368	$\pm$	57		\\
93	&	95437-01-04-01	&	55485.63	&	303	$\pm$	5	&	18	$\pm$	4	&	48	$\pm$	15	&	1.74	$\pm$	0.35	&	5.1	$\pm$	1.0	&	102	$\pm$	21	&	313	$\pm$	89		\\
94	&	95437-01-04-01	&	55485.63	&	356	$\pm$	6	&	17	$\pm$	5	&	36	$\pm$	7	&	1.81	$\pm$	0.25	&	6.3	$\pm$	0.9	&	102	$\pm$	14	&	369	$\pm$	72		\\
95	&	95437-01-04-01	&	55485.63	&	319	$\pm$	6	&	23	$\pm$	3	&	19	$\pm$	4	&	1.74	$\pm$	0.29	&	7.1	$\pm$	1.2	&	78	$\pm$	13	&	432	$\pm$	101		\\
96	&	95437-01-04-01	&	55485.64	&	271	$\pm$	6	&	50	$\pm$	5	&	20	$\pm$	5	&	1.72	$\pm$	0.29	&	6.8	$\pm$	1.1	&	69	$\pm$	12	&	418	$\pm$	100		\\
97	&	95437-01-04-01	&	55485.64	&	332	$\pm$	8	&	24	$\pm$	5	&	19	$\pm$	5	&	1.75	$\pm$	0.37	&	4.0	$\pm$	0.9	&	145	$\pm$	31	&	242	$\pm$	73		\\
98	&	95437-01-04-01	&	55485.64	&	323	$\pm$	6	&	17	$\pm$	3	&	9	$\pm$	3	&	1.81	$\pm$	0.45	&	4.7	$\pm$	1.2	&	126	$\pm$	31	&	273	$\pm$	96		\\
99	&	95437-01-04-01	&	55485.68	&	***			&	40	$\pm$	6	&	15	$\pm$	3	&	1.78	$\pm$	0.39	&	4.4	$\pm$	1.0	&	*		*	&	262	$\pm$	81		\\
100	&	95437-01-04-01	&	55485.68	&	373	$\pm$	6	&	5	$\pm$	1	&	41	$\pm$	9	&	1.69	$\pm$	0.26	&	7.9	$\pm$	1.2	&	79	$\pm$	12	&	497	$\pm$	108		\\
101	&	95437-01-04-01	&	55485.68	&	383	$\pm$	3	&	14	$\pm$	3	&	42	$\pm$	7	&	1.74	$\pm$	0.34	&	6.5	$\pm$	1.3	&	102	$\pm$	20	&	397	$\pm$	111		\\
102	&	95437-01-04-01	&	55485.69	&	263	$\pm$	22	&	110	$\pm$	22	&	45	$\pm$	20	&	1.81	$\pm$	0.27	&	8.9	$\pm$	1.3	&	54	$\pm$	9	&	519	$\pm$	111		\\
103	&	95437-01-04-01	&	55485.69	&	379	$\pm$	22	&	24	$\pm$	5	&	26	$\pm$	9	&	1.68	$\pm$	0.52	&	3.2	$\pm$	1.0	&	200	$\pm$	63	&	201	$\pm$	87		\\
104	&	95437-01-04-01	&	55485.70	&	231	$\pm$	38	&	95	$\pm$	37	&	32	$\pm$	9	&	1.74	$\pm$	0.52	&	3.6	$\pm$	1.1	&	111	$\pm$	38	&	221	$\pm$	94		\\
105	&	95437-01-04-01	&	55485.70	&	337	$\pm$	38	&	81	$\pm$	8	&	34	$\pm$	8	&	1.85	$\pm$	0.82	&	3.3	$\pm$	1.5	&	189	$\pm$	87	&	189	$\pm$	119		\\
106	&	95437-01-04-01	&	55485.70	&	436	$\pm$	8	&	15	$\pm$	3	&	17	$\pm$	3	&	1.86	$\pm$	0.37	&	5.0	$\pm$	1.0	&	161	$\pm$	32	&	287	$\pm$	80		\\
107	&	95437-01-04-01	&	55485.71	&	386	$\pm$	4	&	7	$\pm$	3	&	37	$\pm$	6	&	1.80	$\pm$	0.20	&	6.0	$\pm$	0.7	&	116	$\pm$	13	&	353	$\pm$	55		\\

\end{longtable}	\end{landscape}																												
\end{center}																													
\twocolumn

﻿


\label{lastpage}
\end{document}